\documentclass[preprint,showpacs,aps]{revtex4}
\usepackage{amsmath}
\usepackage{graphicx}

\input{tcilatex}

\begin{document}

\title{Exact Global Solutions of Brane Universes and Big Bounce}
\author{Hongya Liu}
\email{hyliu@dlut.edu.cn}
\affiliation{Department of Physics, Dalian University of Technology, Dalian 116024, P.R.
China}

\begin{abstract}
Exact global solutions of five-dimensional cosmological models compactified
on a $S_{1}$/$Z_{2}$ orbifold with two 3-branes are presented, and evolution
of a simple model is studied. It is found that on all 4D spacetime
hypersurfaces, except a singular one, the expanding universe was started not
from a big bang but from a big bounce, and before the bounce the universe
was in a deflationary contracting phase. It is also found that whether the
energy density on the second brane is positive, zero, or negative depends on
the size of the fifth dimension.

\vspace{1cm}
\end{abstract}

\maketitle

It was proposed [1-3] that our universe is a 3-brane embedded in a higher
dimensional space. While gravity can freely propagate in all dimensions, the
standard matter particles and forces are confined to the 3-brane. Binetruy,
Deffayet and Langlois (BDL) have considered a 5D cosmological model and
derived the Friedmann equations on the branes [4]. This model has received
extensive studies and most of these studies were focused on the branes only
[5]. Be aware that it is technically difficult to move from the branes and
construct exact solutions of the field equations in the bulk, whereas
numerous exact solutions are known in 5D Kaluza-Klein theories. This
suggests that we can take a solution of the Kaluza-Klein theory and use the $
Z_{2}$ reflection symmetry and Israel's jump conditions to obtain a global
brane model. In what follows, we will do this for a rich class of 5D
cosmological solutions found originally by Liu and Mashhoon [6] and
restudied recently by Liu and Wesson [7,8]. We will then construct a class
of global brane solutions which is of the simplest BDL'type, i.e., the
matter on the two branes is a perfect fluid, the bulk is empty, and no
cosmological constants are added in the bulk and on the branes.

The 5D metric for the solutions is 
\begin{equation}
dS^{2}=B^{2}dt^{2}-A^{2}\left( \frac{dr^{2}}{1-kr^{2}}+r^{2}d\Omega
^{2}\right) -dy^{2}.  \label{5metr}
\end{equation}
where $B=B(t_{,}y)$ and $A=A(t_{,}y)$ are two scale factors, $k$($=\pm 1$ or 
$0$) is the 3D curvature index, and $d\Omega ^{2}\equiv (d\theta ^{2}+\sin
^{2}\theta d\phi ^{2})$. A class of general solutions of the 5D vacuum
equations 
\begin{equation}
R_{AB}=0\;,\quad A,B=0123;5\;  \label{Rab=0}
\end{equation}
was given [7] by 
\begin{equation}
A^{2}=\left( \mu ^{2}+k\right) y^{2}+2\nu y+\frac{\nu ^{2}+K}{\mu ^{2}+k}
\;,\qquad B=\frac{1}{\mu }\frac{\partial A}{\partial t}\equiv \frac{\overset{
\mathbf{.}}{A}}{\mu }\;,  \label{5sol}
\end{equation}
where an overdot denotes partial derivative with respect to $t$, $\mu =\mu
(t)$ and $\nu =\nu (t)$ are two arbitrary functions, and $K$ is a 5D
curvature constant related to the square of the Riemann-Christoffel tensor
(Kretschmann scalar) via 
\begin{equation}
R_{ABCD}R^{ABCD}=\frac{72K^{2}}{A^{8}}\;\quad .  \label{RR}
\end{equation}
To confirm solutions (\ref{5sol}), one can go to see the detailed derivation
of the solutions in Ref. [6] (in which different notations were used); or
one can substitute (\ref{5sol}) into $R_{AB}$ directly (which is
complicated); or one can use computer programs such as MAPLE or GRTensor.

Because solutions (\ref{5sol}) satisfy the 5D vacuum equations (\ref{Rab=0}
), we can use them as the bulk solutions of the BDL-type brane models. Note
that one can change $y$ to $-y$ without violating the validity of (\ref{5sol}
) as exact solutions of (\ref{Rab=0}). Therefore, to obtain brane models we
use the $Z_{2}$ reflection symmetry on (\ref{5sol}), i.e., we let 
\begin{equation}
A^{2}=\left( \mu ^{2}+k\right) y^{2}-2\nu \left| y\right| +\frac{\nu ^{2}+K}{
\mu ^{2}+k}\;,\qquad B=\frac{1}{\mu }\frac{\partial A}{\partial t}\equiv 
\frac{\overset{\mathbf{.}}{A}}{\mu }\;.  \label{bran5sol}
\end{equation}
Then we take the corresponding 5D Einstein equations being 
\begin{eqnarray}
G_{AB} &=&k_{\left( 5\right) }^{2}T_{AB}  \notag \\
T_{B}^{A} &=&\delta (y)\text{ diag }\left( \rho
_{1},-p_{1},-p_{1},-p_{1},0\right)  \notag \\
&&+\delta (y-y_{2})\text{ diag }\left( \rho
_{2},-p_{2},-p_{2},-p_{2},0\right) \qquad .  \label{Tab}
\end{eqnarray}
Here the first brane is at $y=y_{1}=0$ and the second is at $y=y_{2}>0$. In
the bulk we have $T_{AB}=0$ and then $G_{AB}=0$, so equations (\ref{Tab})
are satisfied by (\ref{bran5sol}). On the branes we have to solve the
equations (\ref{Tab}) as follows.

We use the 5D metric (\ref{5metr}) to calculate the 5D Einstein tensor,
which, substituted in (\ref{Tab}), gives 
\begin{eqnarray}
G_{00} &=&3\left( \frac{\overset{\mathbf{.}}{A}^{2}}{A^{2}}+k\frac{B^{2}}{
A^{2}}\right) -3B^{2}\left( \frac{A^{\prime \prime }}{A}+\frac{A^{\prime 2}}{
A^{2}}\right)  \notag \\
&=&k_{\left( 5\right) }^{2}B^{2}\left[ \rho _{1}\delta (y)+\rho _{2}\delta
(y-y_{2})\right] \;,  \label{G00}
\end{eqnarray}
\begin{equation}
G_{05}=-3\left( \frac{\overset{\mathbf{.}}{A}^{\prime }}{A}-\frac{\overset{.}
{A}}{A}\frac{B^{\prime }}{B}\right) =0\;,  \label{G05}
\end{equation}
\begin{equation}
G_{55}=-\frac{3}{B^{2}}\left[ \frac{\overset{\mathbf{..}}{A}}{A}+\frac{
\overset{\mathbf{.}}{A}}{A}\left( \frac{\overset{\mathbf{.}}{A}}{A}-\frac{
\overset{\mathbf{.}}{B}}{B}\right) +k\frac{B^{2}}{A^{2}}\right] +\frac{
3A^{\prime }}{A}\left( \frac{A^{\prime }}{A}+\frac{B^{\prime }}{B}\right)
=0\;,  \label{G55}
\end{equation}
\begin{eqnarray}
\left( 1-kr^{2}\right) G_{11} &=&r^{-2}G_{22}=r^{-2}\sin ^{-2}\theta G_{33} 
\notag \\
&=&-\frac{A^{2}}{B^{2}}\left[ \frac{2\overset{\mathbf{..}}{A}}{A}+\frac{
\overset{\mathbf{.}}{A}}{A}\left( \frac{\overset{\mathbf{.}}{A}}{A}-\frac{2
\overset{\mathbf{.}}{B}}{B}\right) +k\frac{B^{2}}{A^{2}}\right]  \notag \\
&&+A^{2}\left[ \frac{B^{\prime \prime }}{B}+\frac{2A^{\prime \prime }}{A}+
\frac{A^{\prime }}{A}\left( \frac{A^{\prime }}{A}+\frac{2B^{\prime }}{B}
\right) \right]  \notag \\
&=&k_{\left( 5\right) }^{2}A^{2}\left[ p_{1}\delta (y)+p_{2}\delta (y-y_{2})
\right] \;.  \label{G11}
\end{eqnarray}
According to Israel's jump conditions, the two scale factors $A$ and $B$ are
required to be continuous across the two branes. Their first derivatives
with respect to $y$ can be discontinuous across the branes, and then their
second derivatives give a Dirac delta function. So we have to calculate $
A^{\prime }$ and $B^{\prime }$ across the two branes. By differentiating $
A^{2}$ in (\ref{bran5sol}) with respect to $y$, we obtain 
\begin{equation}
AA^{\prime }=\left( \mu ^{2}+k\right) y-\nu \frac{\partial \left| y\right| }{
\partial y}\;.  \label{AAprim}
\end{equation}
Therefore we get 
\begin{eqnarray}
A^{\prime }(0^{+}) &=&-\frac{\nu }{A_{1}}\;,\quad A^{\prime }(0^{-})=\frac{
\nu }{A_{1}}\;,  \notag \\
A^{\prime }(y_{2}^{+}) &=&-\frac{\mu ^{2}+k}{A_{2}}y_{2}+\frac{\nu }{A_{2}}
\;,\quad A^{\prime }(y_{2}^{-})=\frac{\mu ^{2}+k}{A_{2}}y_{2}-\frac{\nu }{
A_{2}}\;,  \label{Aprim}
\end{eqnarray}
where $A_{1}\equiv A(t,y=y_{1}=0)$ and $A_{2}\equiv A(t,y=y_{2})$. Then,
using (\ref{Aprim}) and the second equation in (\ref{bran5sol}), i.e., $B=
\overset{\mathbf{.}}{A}/\mu $, we get 
\begin{eqnarray}
B^{\prime }(0^{+}) &=&-\frac{1}{\mu }\frac{\partial }{\partial t}\left( 
\frac{\nu }{A_{1}}\right) \;,\quad B^{\prime }(0^{-})=\frac{1}{\mu }\frac{
\partial }{\partial t}\left( \frac{\nu }{A_{1}}\right) \;,  \notag \\
B^{\prime }(y_{2}^{+}) &=&-\frac{1}{\mu }\frac{\partial }{\partial t}\left( 
\frac{\mu ^{2}+k}{A_{2}}y_{2}-\frac{\nu }{A_{2}}\right) \;,\quad B^{\prime
}(y_{2}^{-})=\frac{1}{\mu }\frac{\partial }{\partial t}\left( \frac{\mu
^{2}+k}{A_{2}}y_{2}-\frac{\nu }{A_{2}}\right) \;.  \label{Bprim}
\end{eqnarray}
So the jumps of $A^{\prime }$ and $B^{\prime }$ across the two branes are 
\begin{eqnarray}
\left[ A^{\prime }\right] _{1} &=&-\frac{2\nu }{A_{1}}\;,\quad \left[
A^{\prime }\right] _{2}=-2\left( \frac{\mu ^{2}+k}{A_{2}}y_{2}-\frac{\nu }{
A_{2}}\right) \;,  \notag \\
\left[ B^{\prime }\right] _{1} &=&-\frac{2}{\mu }\frac{\partial }{\partial t}
\left( \frac{\nu }{A_{1}}\right) \;,\quad \left[ B^{\prime }\right] _{2}=-
\frac{2}{\mu }\frac{\partial }{\partial t}\left( \frac{\mu ^{2}+k}{A_{2}}
y_{2}-\frac{\nu }{A_{2}}\right) \;,  \label{[ABprim]}
\end{eqnarray}
where $\left[ A^{\prime }\right] _{1}\equiv A^{\prime }(0^{+})-A^{\prime
}(0^{-})$ and so on. Substituting (\ref{[ABprim]}) in the field equations (
\ref{G00}) and (\ref{G11}), we obtain 
\begin{eqnarray}
k_{\left( 5\right) }^{2}\,\rho _{1} &=&-\frac{3}{A_{1}}\left[ A^{\prime }
\right] _{1}=\frac{6\nu }{A_{1}^{2}}\;,  \notag \\
k_{\left( 5\right) }^{2}\,p_{1} &=&\frac{1}{B_{1}}\left[ B^{\prime }\right]
_{1}+\frac{2}{A_{1}}\left[ A^{\prime }\right] _{1}=-\frac{2}{\overset{
\mathbf{.}}{A}_{1}}\frac{\partial }{\partial t}\left( \frac{\nu }{A_{1}}
\right) -\frac{4\nu }{A_{1}^{2}}\;,  \label{rho1}
\end{eqnarray}
and 
\begin{eqnarray}
k_{\left( 5\right) }^{2}\,\rho _{2} &=&-\frac{3}{A_{2}}\left[ A^{\prime }
\right] _{2}=\frac{6}{A_{2}}\left( \frac{\mu ^{2}+k}{A_{2}}y_{2}-\frac{\nu }{
A_{2}}\right) \;,  \notag \\
k_{\left( 5\right) }^{2}\,p_{2} &=&\frac{1}{B_{2}}\left[ B^{\prime }\right]
_{2}+\frac{2}{A_{2}}\left[ A^{\prime }\right] _{2}  \notag \\
&=&-\frac{2}{\overset{\mathbf{.}}{A}_{2}}\frac{\partial }{\partial t}\left( 
\frac{\mu ^{2}+k}{A_{2}}y_{2}-\frac{\nu }{A_{2}}\right) -\frac{4}{A_{2}}
\left( \frac{\mu ^{2}+k}{A_{2}}y_{2}-\frac{\nu }{A_{2}}\right) \;.
\label{rho2}
\end{eqnarray}
Thus we have derived the energy densities and pressures on the two branes.
Meanwhile, the conservation law $T_{A\;;B}^{\;B}=0$ gives 
\begin{equation}
\overset{.}{\rho }_{i}+3\left( \rho _{i}+p_{i}\right) \frac{\overset{\mathbf{
.}}{A}_{i}}{A_{i}}=0\;,\quad i=1,2\;.  \label{conserv}
\end{equation}
This relation can also be verified directly by substituting (\ref{rho1}), (
\ref{rho2}) and (\ref{bran5sol}) into (\ref{conserv}) as expected.

From the 5D metric (\ref{5metr}) we see that on a given $y=$ constant 4D
hypersurface the proper time can be defined as $d\tau =B(t,y)dt$. So the
Hubble and deceleration parameters can be defined as 
\begin{equation}
H(t,y)\equiv \frac{1}{B}\frac{\overset{\mathbf{.}}{A}}{A}=\frac{\mu }{A}
\;,\qquad q(t,y)\equiv -\frac{A}{B}\frac{\partial }{\partial t}\left( \frac{
\overset{\mathbf{.}}{A}}{B}\right) \left/ \left( \frac{\overset{\mathbf{.}}{A
}}{B}\right) ^{2}\right. =-\frac{A\overset{\mathbf{.}}{\mu }}{\mu \overset{
\mathbf{.}}{A}}  \label{H,q}
\end{equation}
where we have used the relation $B=\overset{\mathbf{.}}{A}/\mu $. Meanwhile,
the first equation in (\ref{bran5sol}) for the branes can be written as 
\begin{equation}
(\mu ^{2}+k)A_{i}^{2}=\left[ (\mu ^{2}+k)y_{i}-\nu \right] ^{2}+K\;,\quad
i=1,2\;.  \label{A^2i}
\end{equation}
Then, using (\ref{rho1}), (\ref{rho2}) and (\ref{H,q}) in (\ref{A^2i}), we
obtain 
\begin{equation}
H_{i}^{2}+\frac{k}{A_{i}^{2}}=\frac{k_{\left( 5\right) }^{4}}{36}\rho
_{i}^{2}+\frac{K}{A_{i}^{4}}\;,\quad i=1,2\;,  \label{Hi-rho}
\end{equation}
\begin{equation}
H_{i}^{2}\left( 1-2q_{i}\right) +\frac{k}{A_{i}^{2}}=-\frac{k_{\left(
5\right) }^{4}}{12}\rho _{i}\left( \rho _{i}+2p_{i}\right) -\frac{K}{
A_{i}^{4}}\;,\quad i=1,2\;.  \label{Hi-pi}
\end{equation}
These two equations are the induced Friedmann equations on the branes.

Thus we obtain a complete set of global exact solutions given in (\ref
{bran5sol}) and (\ref{rho1})-(\ref{Hi-pi}) which contains two arbitrary
functions $\mu (t)$ and $\nu (t)$. From the relation $B=\overset{\mathbf{.}}{
A}/\mu $ and metric (\ref{5metr}) we see that the form of $Bdt$ is invariant
under an arbitrary coordinate transformation $t\rightarrow \widetilde{t}(t)$
.. This freedom can be used to fix one of the two functions $\mu (t)$ and $
\nu (t)$. Another freedom corresponds, as is in the standard general
relativity, to the unspecified equation of state of matter. For a given
equation of state $p=p(\rho )$, the two equations in (\ref{rho1}) give a
constraint that can be used to determine the function, say, $\nu (t)$. So,
generally speaking, if the matter content on the first brane is known, then $
\mu (t)$ and $\nu (t)$ can be fixed. Then the whole solutions can be fixed
too. Then, by (\ref{rho2}), if the size $y_{2}$ of the fifth dimension is
also known, we will know the matter content on the second brane, and this is
of great importance.

Using above procedure, we can, in principle, construct particular models
such as the matter and radiation dominated models as in the standard FRW
cosmology. However, preliminary studies show that they are not
mathematically simple. So we leave this study in the future. Here, in this
letter, we will use another way to obtain an explicit universe model, that
is, we will choose $\mu (t)$ and $\nu (t)$ firstly\ and then to study the
evolution and matter properties. Now we let 
\begin{equation}
k=0\;,\qquad K=\frac{1}{L^{2}}\;,\qquad \nu (t)=\frac{t_{b}}{Lt}\;,\qquad
\mu (t)=\left( 2Lt\right) ^{-1/2}\quad ,  \label{special}
\end{equation}
where $L$ is a constant with a dimension of length and $t_{b}$ (we assume $
t_{b}>0$) is a critical constant with a dimension of time. In this way the
solutions (\ref{bran5sol}) become 
\begin{eqnarray}
A^{2} &=&\frac{2t}{L}\left[ 1+\left( \frac{\left| y\right| -2t_{b}}{2t}
\right) ^{2}\right] \;,  \notag \\
B^{2} &=&\left[ 1-\left( \frac{\left| y\right| -2t_{b}}{2t}\right) ^{2}
\right] ^{2}\left/ \left[ 1+\left( \frac{\left| y\right| -2t_{b}}{2t}\right)
^{2}\right] \right. \;.  \label{ABspec}
\end{eqnarray}
So on the first brane we have 
\begin{eqnarray}
A_{1}^{2} &=&\frac{2t}{L}\left[ 1+\left( \frac{t_{b}}{t}\right) ^{2}\right]
\;,\qquad B_{1}^{2}=\left[ 1-\left( \frac{t_{b}}{t}\right) ^{2}\right] ^{2}
\left[ 1+\left( \frac{t_{b}}{t}\right) ^{2}\right] ^{-1}\;,  \notag \\
k_{\left( 5\right) }^{2}\rho _{1} &=&\frac{3t_{b}}{t^{2}+t_{b}^{2}}\;,\qquad
k_{\left( 5\right) }^{2}p_{1}=\frac{2t_{b}}{t^{2}-t_{b}^{2}}-\frac{t_{b}}{
t^{2}+t_{b}^{2}}\;.  \label{Bran1spec}
\end{eqnarray}
On the second brane we have 
\begin{eqnarray}
A_{2}^{2} &=&\frac{2t}{L}\left[ 1+\left( \frac{y_{2}-2t_{b}}{2t}\right) ^{2}
\right] \;,\quad B_{2}^{2}=\left[ 1-\left( \frac{y_{2}-2t_{b}}{2t}\right)
^{2}\right] ^{2}\left[ 1+\left( \frac{y_{2}-2t_{b}}{2t}\right) ^{2}\right]
^{-1}\;  \notag \\
k_{\left( 5\right) }^{2}\,\rho _{2} &=&\frac{6\left( y_{2}-2t_{b}\right) }{
4t^{2}+\left( y_{2}-2t_{b}\right) ^{2}}\;,\quad k_{\left( 5\right)
}^{2}\,p_{2}=\frac{4\left( y_{2}-2t_{b}\right) }{4t^{2}-\left(
y_{2}-2t_{b}\right) ^{2}}-\frac{2\left( y_{2}-2t_{b}\right) }{4t^{2}+\left(
y_{2}-2t_{b}\right) ^{2}}\;.  \label{Bran2spec}
\end{eqnarray}
Equations (\ref{ABspec})-(\ref{Bran2spec}) constitute a quite simple
two-brane model.

From (\ref{ABspec}) we see that, in a $y=$ constant hypersurface, the scale
factors $B\approx 1$ and $A\approx \sqrt{2t/L}$ for $2t\gg \left| \left|
y\right| -2t_{b}\right| $. So it approaches the standard radiation-dominated
model at late times of the universe. The global evolution of the scale
factor $A(t,y)$ with $t_{b}=1$ and $L=1$ is plotted in Fig.1. From this
figure we see that there is a singular surface $\left| y\right| =2t_{b}$ in
which $A\rightarrow 0$ as $t\rightarrow 0$ , showing a \textbf{big bang}
singularity as is in the standard FRW models. However, in all other
hypersurfaces, the scale factor $A$ reaches a \textbf{non-zero minimum} $
\sqrt{2t/L}$ at $2t=\left| \left| y\right| -2t_{b}\right| $ (where $B=0$);
and at both sides of this minimum, $A$ tends to infinity. So this minimum
can naturally be explained as a \textbf{big bounce }[7]. Generally, the
solution has two kinds of singularities, corresponding to $A=0$ and $B=0$
respectively as discussed in Ref. [7]. $A=0$ is, by (\ref{RR}), an intrinsic
singularity of the 5D manifold, whereas $B=0$ is just a coordinate
singularity similar to the Schwarzschild event horizon. So here the big
bounce singularity belongs to the second kind. 
\begin{figure}[tbp]
\centering\includegraphics[width=2.65in,height=2.65in]{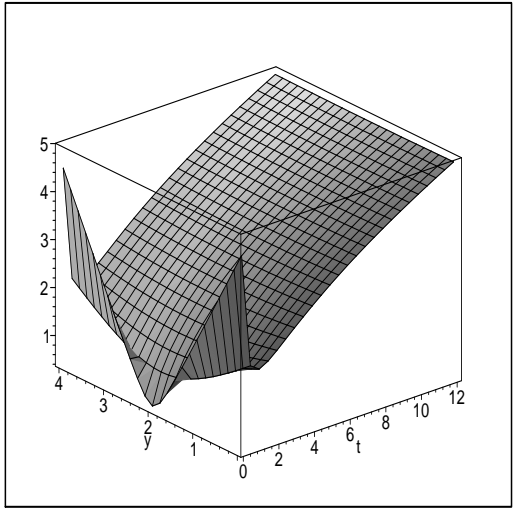}
\caption{Global evolution of the scale factor $A(t,y)=\protect\sqrt{
2t+\left( \left| y\right| -2\right) ^{2}/(2t)}$. The first brane (our
universe) is at $y=0$; while the second brane (the hidden universe) could be
placed at $y=2$, $\;0<y<2$, or $y>2$, giving three types of brane
cosmological models, respectively.}
\end{figure}

The first brane is at $y=0$. By (\ref{Bran1spec}) we see that at the bounce
point $t=t_{b}$ we have $A_{1}$ tends to its minimum, $B_{1}$ tends to zero, 
$\rho _{1}$ is finite, and $p_{1}$ tends to infinity. Physically, we can say
that it is this infinitely large pressure $p_{1}$ that caused the big bounce.

From (\ref{Bran2spec}) we see that the evolution and the matter properties
on the second brane depend on the size $y_{2}$ of the fifth dimension (see
Fig. 1 also). Thus we have three types of brane models as shown below.

\begin{description}
\item \textbf{Type I}. The second brane is at $y_{2}=2t_{b}$. Then (\ref
{Bran2spec}) gives $A_{2}=\sqrt{2t/L}$, $B_{2}=1$, and$\quad \rho
_{2}=p_{2}=0$. This implies that the second brane is empty of matter. So
Type I is actually a \textbf{one-brane model} for which there was a big
bounce on our side of the model and a big bang on the ``hidden'' side.

\item \textbf{Type II}. The second brane is at $0<y_{2}<2t_{b}$. Then (\ref
{Bran2spec}) shows that there were bounces on both branes. On the second
brane we have $\rho _{2}<0$. So Type II is a \textbf{two-brane model} for
which the energy density on the second brane is \textbf{negative}.

\item \textbf{Type III}. The second brane is at $y_{2}>2t_{b}$. This is also
a \textbf{two-brane model} for which the energy density on the second brane
is \textbf{positive}. Specifically, if $y_{2}=4t_{b}$, then $A_{2}=A_{1}$, $
B_{2}=B_{1}$, $\rho _{2}=\rho _{1}$, and $p_{2}=p_{1}$, giving a completely
symmetric two-brane model.
\end{description}

Now let us consider the solution (\ref{Bran1spec}) for which when the
coordinate time $t$ varies from $0$ to $t_{b}$ and then to $+\infty $, the
spatial scale factor $A_{1}(t)$ contracts from $+\infty $ to a minimum $
A_{\min }=2\sqrt{t_{b}/L}>0$ and then expands to $+\infty $ again. We also
find that as $t\rightarrow 0$, $A_{1}^{2}\rightarrow 2t_{b}^{2}/(Lt)$, $
B_{1}^{2}\rightarrow \left( t_{b}/t\right) ^{2}$ and the 4D metric on the
first brane tends to 
\begin{equation}
ds_{1}^{2}\rightarrow \left( \frac{t_{b}}{t}\right) ^{2}dt^{2}-\frac{
2t_{b}^{2}}{Lt}\left( dr^{2}+r^{2}d\Omega ^{2}\right) \quad \text{(as }
t\rightarrow 0\text{)}\;.  \label{ds1-spec}
\end{equation}
By a coordinate transformation 
\begin{equation}
t=2L^{-1}t_{b}^{2}e^{\tau /t_{b}}\;,  \label{t-trans}
\end{equation}
we get 
\begin{equation}
ds_{1}^{2}\rightarrow d\tau ^{2}-e^{-\frac{\tau }{t_{b}}}\left(
dr^{2}+r^{2}d\Omega ^{2}\right) \quad \text{(as }\tau \rightarrow -\infty 
\text{)}\;,  \label{ds1-spec2}
\end{equation}
which represents a \textit{deflationary} de Sitter cosmological model. Note
that by (\ref{t-trans}) $t=0$ corresponds to $\tau =-\infty $. Therefore we
conclude that according to the proper time $\tau $, the universe on the
first brane has been existed forever and contracts from a de Sitter space to
a non-zero minimum, at which the pressure $p_{1}$ reaches to infinity and
causes a big bounce. After then, the universe expands.

In conclusion, we have derived a class of exact global solutions of
five-dimensional cosmological models with two 3-branes, for which matter on
the branes is of the form of a perfect fluid, the bulk is empty, and no
cosmological constants were introduced. This solutions contain two arbitrary
constants $k$ and $K$, corresponding to the 3D and 5D curvatures
respectively, and two arbitrary functions $\mu (t)$ and $\nu (t)$,
corresponding to the two freedoms: the arbitrary coordinate transformation $
t\rightarrow \widetilde{t}(t)$\ and the unspecified equation of state of
matter. By choosing $\mu (t)$ and $\nu (t)$ properly a special model is
discussed in detail. It is found that in the late times of the universe this
model evolves at the same rate as is in the standard radiation-dominated FRW
model. And there is a singular 4D hypersurface $y=0$ on which there was a
big bang. On all other hypersurfaces there were no big bang but big bounces.
Before the bounces, the universe was in a contracting phase. It is also
found that whether the energy density on the second brane is positive, zero,
or negative depend on the size of the fifth dimension. Thus three types of
models are obtained for which type I is actually a one-brane model while
types II and III are two-brane models with negative and positive energy
densities, respectively.

{\Large ACKNOWLEDGMENTS\medskip }

We thank Paul Wesson and Guowen Peng for comments. This work was supported
by NSF of P. R. China under Grants 19975007 and 10273004.


\begin{thebibliography}{9}
\bibitem{ADD} \baselineskip 24ptN. Arkani-Hamed, S. Dimopoulos and G. Dvali,
Phys. Lett. B\textbf{\ 429}, 263 (1998), hep-ph/9803315; Phys. Rev. D 
\textbf{59}, 086004 (1999), hep-ph/9807344; I. Antoniadis, N. Arkani-Hamed,
S. Dimopoulos and G. Dvali, Phys. Lett. B\textbf{\ 436}, 257 (1998),
hep-ph/9804398.

\bibitem{HW} P. Horava and E. Witten, Nucl. Phys. B \textbf{460}, 506
(1996), hep-th/9510209; E. Witten, Nucl. Phys. B \textbf{471}, 135 (1996),
hep-th/9602070; P. Horava and E. Witten, Nucl. Phys. B. \textbf{475}, 94
(1996), hep-th/9603142.

\bibitem{RR} L. Randall and R. Sundrum, Phys. Rev. Lett. \textbf{83}, 3370
(1999), hep-ph/9905221; Phys. Rev. Lett. \textbf{83}, 4690 (1999),
hep-th/9906064.

\bibitem{BDL} P. Binetruy, C. Deffayet and D. Langlois, Nucl. Phys. B 
\textbf{565}, 269 (2000), hep-th/9905012.

\bibitem{Cline99} See, for example, J.M. Cline, C. Grojean, and G. Servant,
Phys. Rev. Lett. \textbf{83}, 4245 (1999), hep-ph/9906523; C. Csaki, M.
Graesser, C. Kolda, and J. Terning, Phys. Lett. B \textbf{462}, 34 (1999),
hep-ph/9906513; P. Kanti, I. Kogan, K.A. Olive and M. Pospelov, Phys. Lett.
B \textbf{468}, 31 (1999), hep-ph/9909481; Phys. Rev. D \textbf{61}, 106004
(2000), hep-ph/9912266; S. Mukohyama, T. Shiromizu and K. Maeda, Phys. Rev.
D \textbf{62}, 024028 (2000), hep-th/9912287; S. Mukohyama, Phys. Lett. B 
\textbf{473}, 241 (2000), hep-th/9911165; J. Khoury and R.-J. Zhang, Phys.
Rev. Lett. \textbf{89}, 061302 (2002), hep-th/0203274.

\bibitem{L-M95} Hongya Liu and B. Mashhoon, Ann. Phys. (Leipzig) \textbf{4},
565 (1995).

\bibitem{LWapj01} Hongya Liu and P.S. Wesson, Astrophys. J. \textbf{562}, 1
(2001), gr-qc/0107093.

\bibitem{Leon01} J. Ponce de Leon, Mod. Phys. Lett. A. \textbf{16}, 2291
(2001), gr-qc/0111011.
\end{thebibliography}
\end{document}